\documentclass[a4paper,11pt]{article}
\usepackage{pos}
\usepackage{subfigure}

\title{The Thirring Model in 2+1$d$ with Optimised Domain Wall Fermions}

\author*[a]{Simon Hands}
\author[b]{Jude Worthy}

\affiliation[a]{Department of Mathematical Sciences, University of Liverpool,\\
Liverpool L69 3BX, United Kingdom}

\affiliation[b]{Department of Physics, Swansea University,\\
Singleton Park, Swansea SA2 8PP, United Kingdom}

\emailAdd{Simon.Hands@liverpool.ac.uk}
\emailAdd{judeworthy@gmail.com}

\abstract{After briefly reviewing the potential for the $N$-flavor Thirring
model, formulated with reducible fermions in 2+1$d$, to exhibit a
strongly-coupled UV-stable fixed point where U($2N$) symmetry is spontaneously broken
by a fermion bilinear condensate, we present recent lattice studies
using the Domain Wall Fermion formulation. In particular, we focus on possible
improved methods  for  extracting the necessary $L_s\to\infty$ limit, where $L_s$ is the wall
separation, through a combination of partial
quenching (ie. $L_s({\rm valence})>L_s({\rm sea})$), replacing the Shamir kernel
with the Wilson kernel in the definition of the overlap operator, and improved
estimation of the signum function using the Zolotarev approximation. Equation of
state fits  for critical exponents 
on $12^3$ systems yield encouraging agreement between
distinct  approaches, consistent with universal scaling, while contradicting earlier fits based on a naive
extrapolation. The new results are  also in tension with old results obtained with
staggered fermions.}

\FullConference{The 40th International Symposium on Lattice Field Theory (Lattice 2023)\\
July 31st - August 4th, 2023\\
Fermi National Accelerator Laboratory\\}

\begin{document}
\maketitle

\section{Introduction}
The  Thirring model is a covariant quantum field theory of interacting fermions
with Lagrangian density
\begin{equation}
{\cal
L}=\bar\psi_i(\partial\!\!\!/\,+m)\psi_i+{{g^2}\over{2N}}(\bar\psi_i\gamma_\mu\psi_i)^2.
\label{eq:L}
\end{equation}
Here the index $i$ runs over $N$ flavors. The contact interaction between
currents is repulsive between like charges and attractive between opposite. In
2+1$d$ we may specify the fields $\psi,\bar\psi$ to lie in reducible
representations of the spinor algebra, so that the Dirac matrices $\gamma_\mu$
are $4\times4$, and there is a matrix
$\gamma_5=\gamma_0\gamma_1\gamma_2\gamma_3$ such that
$\{\gamma_5,\gamma_\mu\}=0$. For sufficiently large interaction strength $g^2$
and sufficiently small $N$ the Fock vacuum may be disrupted
through formation of a particle -- antiparticle bilinear condensate 
\begin{equation}
\langle\bar\psi\psi\rangle\equiv{\partial\ln Z\over\partial m}\not=0.
\end{equation}
This results in a dynamically-generated mass gap at Dirac points where
$E(\vec p)=0$, in close analogy to chiral symmetry breaking in QCD. It has been
hypothesised~\cite{Son:2007ja} 
that the transition to non-vanishing condensate at $g_c^2(N)$ defines a Quantum Critical
Point whose universal properties could perhaps characterise low energy
electronic excitations in a planar material such as graphene. Such a fixed point
would correspond to a strongly-interacting quantum field theory with {\it a
priori\/} no
small dimensionless parameters.

In the absence of a bare fermion mass the Lagrangian (\ref{eq:L}) is invariant
under a global U($2N$) generated by the following rotations:
\begin{eqnarray}
\psi\mapsto e^{i\alpha}\psi\;\;\;\bar\psi\mapsto\bar\psi e^{-i\alpha};&\;&\;\;
\psi\mapsto e^{\alpha\gamma_3\gamma_5}\psi,\;\;\;\bar\psi\mapsto\bar\psi
e^{-\alpha\gamma_3\gamma_5};\label{eq:unbroken}\\
\psi\mapsto e^{i\alpha\gamma_3}\psi,\;\;\;\bar\psi\mapsto\bar\psi
e^{i\alpha\gamma_3};&\;&\;\;
\psi\mapsto e^{i\alpha\gamma_5}\psi,\;\;\;\bar\psi\mapsto\bar\psi
e^{i\alpha\gamma_5}.\label{eq:broken}
\end{eqnarray}
Once $m\not=0$ (\ref{eq:broken}) are no longer symmetries, so bilinear
condensation results in a symmetry breaking U($2N)\to\,$U($N)\otimes$U($N$); this
should be compared to the pattern U($N)\otimes$U($N)\to$U($N$) which pertains
either to models built using staggered lattice fermions or continuum models
using the K\"ahler-Dirac formulation of relativistic fermions~\cite{Hands:2021mrg}.

\section{Domain Wall Fermions}
In a series of papers~\cite{Hands:2016foa,Hands:2018vrd,Hands:2020itv} 
we have studied the strong dynamics of this
proposed symmetry breaking through lattice simulations with reducible fermions
implemented through a domain wall construction of the form ${\cal
L}_{\rm kin}=\bar\Psi(x,s)D_{DWF}\Psi(y,s^\prime)$, where $s, s^\prime$ are coordinates
along a fictitious third spatial direction, with open
boundaries (ie. domain walls) separated by distance $L_s$.
As $L_s\to\infty$, near zero-modes of $D_{DWF}$ are localised on the walls as
$\pm$ eigenmodes of $\gamma_3$, and U($2N$)-symmetric 2+1$d$ physics described
in terms of 
\begin{equation}
\psi(x)={\cal P}_-\Psi(x,1)+{\cal P}_+ \Psi(x,L_s);\;\;\;
\bar\psi(x)=\bar\Psi(x,L_s){\cal P}_-+\bar\Psi(x,1){\cal P}_+,
\end{equation}
with projectors ${\cal P}_\pm={1\over2}(1\pm\gamma_3)$.

Now, for an arbitrary Dirac kernel $D$ operating in the target 2+1$d$ space,
the closest we can get to U($2N$) symmetry is articulated by the
Ginsparg-Wilson (GW) relations
\begin{equation}
\left\{\gamma_3,D\right\}=2D\gamma_3D;\;\;
\left\{\gamma_5,D\right\}=2D\gamma_5D;\;\;
[\gamma_3\gamma_5,D]=0.\label{eq:GW2+1}
\end{equation}
By construction (\ref{eq:GW2+1}) are satisfied by the 2+1$d$ overlap operator
\begin{equation}
D_{\rm ov}={1\over2}\left[(1+m)+(1-m){{\cal A}\over{\sqrt{{\cal A}^\dagger{\cal
A}}}}\right];\label{eq:overlap}
\end{equation}
for the choice of {\em Shamir kernel}
\begin{equation}
{\cal A}=[2+D_W-M]^{-1}[D_W-M],
\label{eq:Shamir}
\end{equation}
with $D_W$ the 2+1$d$ Wilson fermion kernel and $Ma=O(1)$ the domain wall
height, the overlap operator $D_{\rm ov}$ can be shown to be equivalent to the $L_s\to\infty$
limit of $D_{DWF}$ used to date~\cite{Hands:2015dyp}:
\begin{equation}
\lim_{L_s\to\infty}{{{\rm det}D_{DWF}(m)}\over{{\rm det}D_{DWF}(m=1)}}={\rm
det}D_{\rm ov}(m).
\end{equation}
\begin{figure}[ht]
\centerline{
  \subfigure[$\Delta(\beta,m)$ on $16^3\times L_s$ \cite{Hands:2020itv}]
     {\includegraphics[width=2.5in]{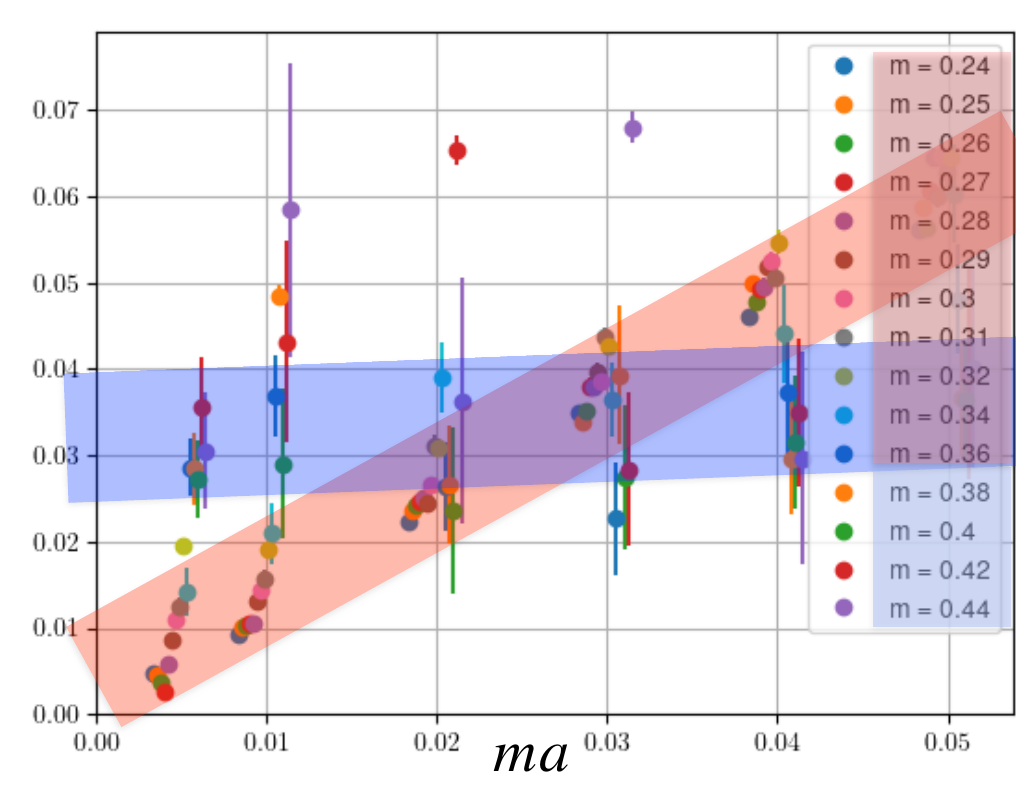}\label{fig:Delta}}
  \hspace*{4pt}
  \subfigure[$\delta_h(\beta,m)$ on $16^3\times48$  \cite{Hands:2020itv}]
     {\includegraphics[width=3.0in]{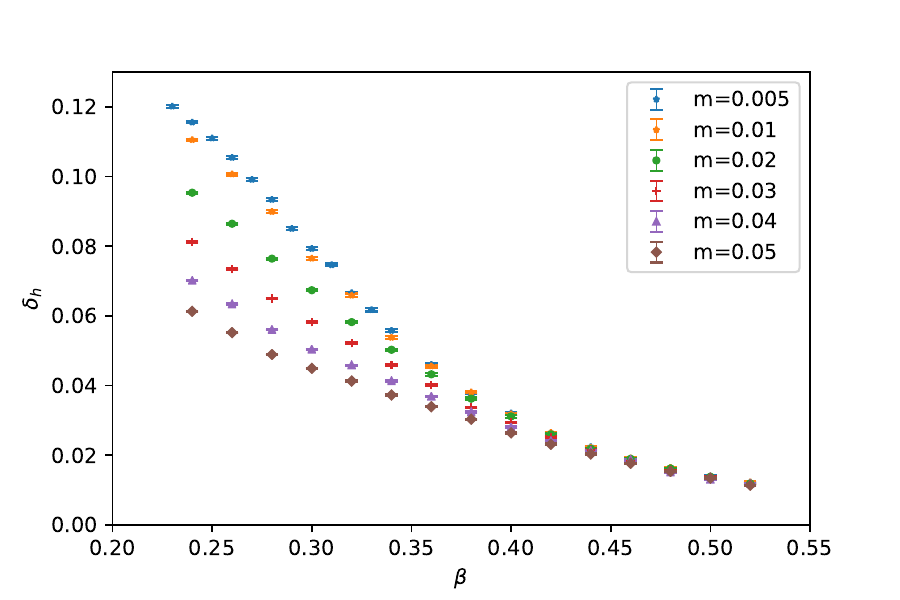}\label{fig:deltapeek}}
}
\caption{Quantifying the approach to $L_s\to\infty$}
\label{fig:Lsextr} 
\end{figure}
The four-fermion interaction in the Thirring model may be reproduced by the
introduction of a bosonic vector auxiliary field $A_\mu$ coupled to the
conserved current $i\bar\psi\gamma_\mu\psi$. In the lattice model set out in 
\cite{Hands:2016foa,Hands:2018vrd,Hands:2020itv}, $A_\mu$ is located on a link of
the 2+1$d$ lattice and linearly coupled to the conserved DWF current which is defined
throughout the bulk. Hence off-diagonal elements of $D_{DWF}$ are of the form
$D_\mu\sim(1+iA_\mu)$ rather than the canonical $D_\mu\sim e^{iA_\mu}$ of
abelian gauge theories: the link fields are thus non-compact and non-unitary, which
makes inversion of $D_{DWF}$ numerically challenging. We simulate the Thirring
model with $N=1$ using the RHMC algorithm to reproduce the functional measure 
$\sqrt{{\rm det}D_{DWF}^\dagger D_{DWF}}$~\cite{Hands:2018vrd}.
Taking $L_s\to\infty$ is hard; we have fitted data
from $16^3\times L_s=8,16,\ldots,80$ using an exponential {\em
Ansatz}
\begin{equation}
\langle\bar\psi\psi\rangle_\infty-\langle\bar\psi\psi\rangle_{L_s}=A(\beta,m)e^{-\Delta(\beta,m)L_s},
\label{eq:Ansatz}
\end{equation}
with inverse coupling $\beta\equiv ag^{-2}$.
Fig.~\ref{fig:Delta} shows a compendium of fitted values for $\Delta$. For weak
coupling $\beta>0.4$, $\Delta$ is roughly $m$-independent (blue band), but for stronger
couplings $\beta<0.35$, $\Delta\propto m$ (red band), implying that here the large-$L_s$ limit
is extremely challenging in the massless limit.
Fig.~\ref{fig:deltapeek} shows a compendium of the residual $\delta_h$ defined
by
\begin{equation}
\delta_h\equiv\Im\langle\bar\Psi(1)\gamma_3\Psi(L_s)\rangle\approx{1\over2}
\left(\langle\bar\psi\psi\rangle-i\langle
\bar\psi\gamma_3\psi\rangle\right)
\end{equation}
which should vanish if the U($2N$) symmetry relating the two condensates on the
RHS is restored. Again, at fixed $L_s$ and strong coupling the symmetry restoration
becomes harder as $m\to0$. Further results for the locality
of $D_{\rm ov}$ and the restoration of the GW relations
(\ref{eq:GW2+1}) can be found in \cite{Hands:2020itv}.

\section{Improving $L_s\to\infty$}
Since lack of control of the $L_s\to\infty$ limit casts doubt on the accuracy 
of earlier studies~\cite{Hands:2018vrd, Hands:2020itv},
we have recently experimented with three strategies for ameliorating the
problem. Further discussion can be found in \cite{Worthy:2021ddb}.

\begin{itemize}

\item
\begin{figure}
\centerline{\includegraphics[width=7.8cm]{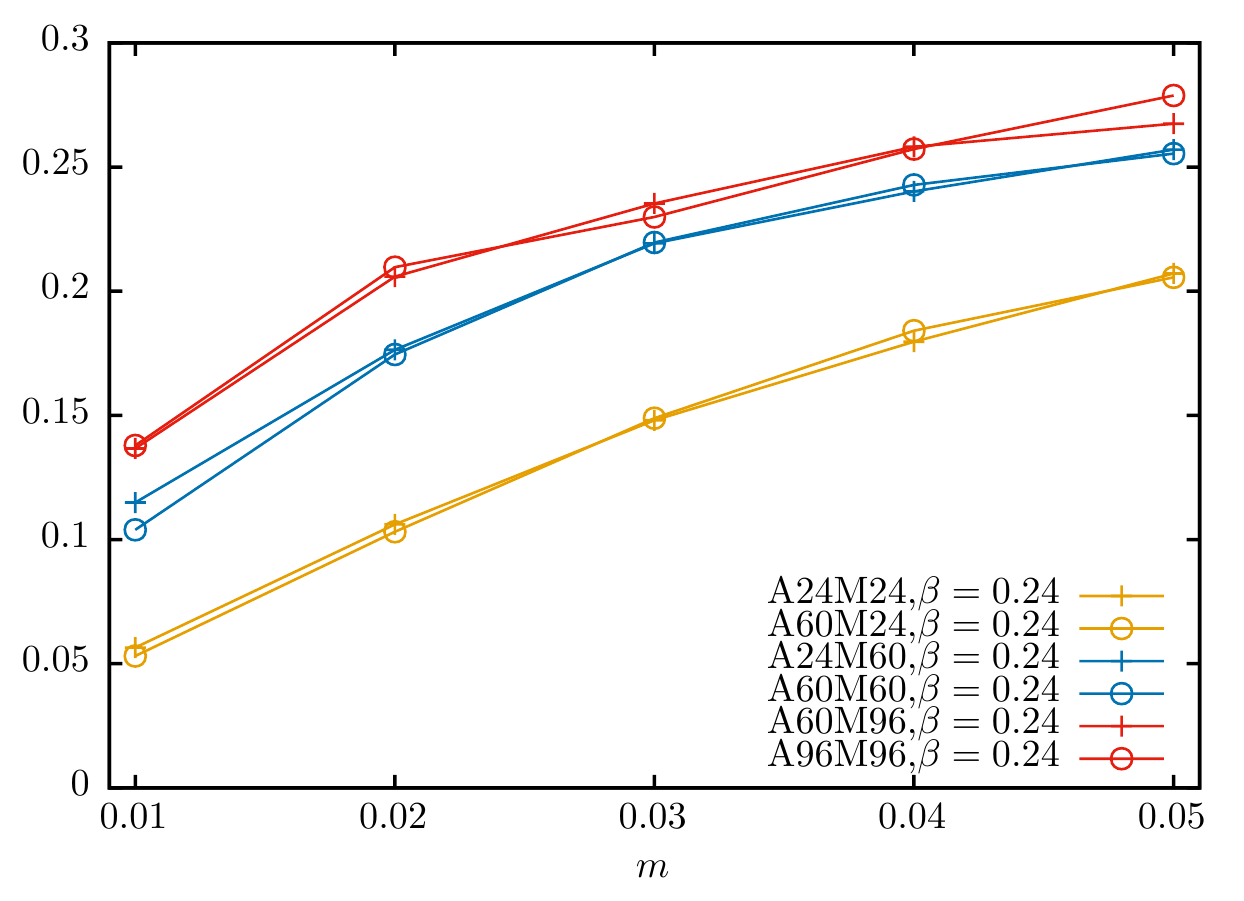}}
\caption{Bilinear condensate for Shamir kernel on $12^3\times L_s$ at $\beta=0.24$,
for various  different $L_s({\rm sea})$ (labelled A in key), for $L_s({\rm valence})=24$ (yellow), 60
(blue), 96 (orange) (labelled M in key).}
\label{fig:PQ}
\end{figure}
{\bf Partial Quenching (PQ):} This is the most straightforward to implement. 
The main impact of finite $L_s$ appears in measurements in
the fermion sector (the chief example being the bilinear condensate order
parameter itself), while the effect on the underlying bosonic $A_\mu$
configurations is much milder, as exemplified in Fig.~\ref{fig:PQ}. 
Accordingly we have made studies with
$L_s({\rm sea})\not=L_s({\rm valence})$,
the most straightforward choice being $L_s({\rm sea})\ll L_s({\rm valence})$.

\item
{\bf Wilson kernel:} We have replaced the Shamir kernel (\ref{eq:Shamir}) in the definition
(\ref{eq:overlap}) of 
$D_{\rm ov}$,  where the corresponding DWF operator is represented as an $L_s\times L_s$
matrix\footnote{The corner elements in (\ref{eq:DSHT},\ref{eq:DWHT}) are
appropriate for a mass term $im\bar\psi\gamma_3\psi$, equivalent to
$m\bar\psi\psi$ after U($2N$) rotation.}
\begin{equation}
D_{SHT}=\left(\begin{matrix}
D_W-M+I&-{\cal P}_- &0    &im{\cal P}_+\\
-{\cal P}_+ &D_W-M+I&-{\cal P}_- &0\\
0    &-{\cal P}_+ &D_W-M+I&-{\cal P}_-\\
-im{\cal P}_- &0    &-{\cal P}_+ &D_W-M+I\\
\end{matrix}\right)\label{eq:DSHT}
\end{equation}
with the {\em Wilson kernel}  ${\cal A}=D_W-M$, with DWF transcription 
\begin{equation}
D_{WHT}=\left(\begin{matrix}
{\cal A}+I&({\cal A}-I){\cal P}_- &0    &-im({\cal A}-I){\cal P}_+\\
({\cal A}-I){\cal P}_+ &{\cal A}+I&({\cal A}-I){\cal P}_- &0\\
0    &({\cal A}-I){\cal P}_+ &{\cal A}+I&({\cal A}-I){\cal P}_-\\
+im({\cal A}-I){\cal P}_- &0    &({\cal A}-I){\cal P}_+ &{\cal A}+I\\
\end{matrix}\right)\label{eq:DWHT}
\end{equation}
which is
much better-conditioned.

\item
\begin{figure}
\centerline{\includegraphics[width=7.8cm]{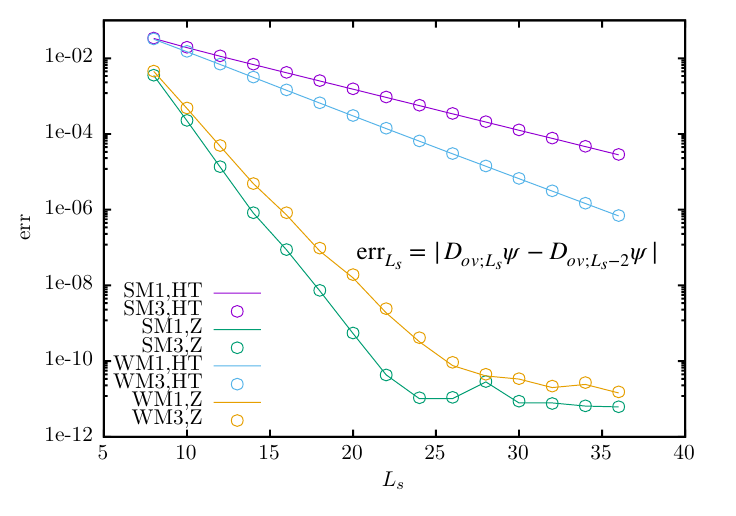}}
\caption{Convergence of the overlap with increasing $L_s$ on a fixed $12^3$ auxiliary
background, for both Shamir and Wilson kernels, and mass terms $m_1\bar\psi\psi,
m_3\bar\psi i\gamma_3\psi$, with $m_1a=m_3a=0.05$.}
\label{fig:HTvsZ}
\end{figure}
{\bf Improved Rational approximation for ${\rm sgn}$:} The approach of $D_{DWF}$
to $D_{\rm ov}$ at finite $L_s$ depends on a rational approximation to ${\rm
sgn}({\cal A})\equiv{\cal A}/\sqrt{{\cal A}^\dagger{\cal A}}$ expressed as a product of $L_s$ factors. We have replaced the
hyperbolic tangent (HT) form 
\begin{equation}
{\rm sgn}(x)\approx\tanh(L_s\tanh^{-1}x)={{1-{\cal T}_{HT}}\over{1+{\cal T}_{HT}}}
\;\;\;{\rm with}\;\;\;
{\cal T}_{HT}=\left({{1-x}\over{1+x}}\right)^{L_s}
\end{equation}
used in vanilla DWF~\cite{Hands:2015dyp} by the Zolotarev (Z)
approximation
\begin{equation}
{\rm sgn}(x)\approx{{1-{\cal T}_{Z}}\over{1+{\cal T}_{Z}}}\equiv
dx{{\prod_{m=1}^{L_s/2-1}(a_m-x^2)}\over{\prod_{m=1}^{L_s/2}(d_m-x^2)}}
\;\;\;{\rm with}\;\;\;
{\cal T}_{Z}=\prod_{s=1}^{L_s}{{1-\omega_sx}\over{1+\omega_sx}},
\end{equation}
where the coefficients $a_m,d_m,d$ depend on the applicable range of the
approximation, chosen to match the spectral range of ${\cal
A}$~\cite{Chiu:2002ir}.
The superior $L_s$-convergence of Z over HT is shown in Fig.~\ref{fig:HTvsZ}.
The coefficients $\omega_s$ found via the roots of ${\rm sgn}(x)=1$ can be used
to replace all instances of ${\cal A}$ in the $s$th row of (\ref{eq:DWHT}) by
$\omega_s{\cal A}$ to yield 
the {\em optimised} DWF introduced by
Chiu~\cite{Chiu:2002ir}.

\end{itemize}

\section{Results for the Equation of State}
\begin{figure}[ht]
\centerline{
  \subfigure[Shamir kernel, $L_s({\rm sea})=30$, $L_s({\rm valence})=300$]
     {\includegraphics[width=2.85in]{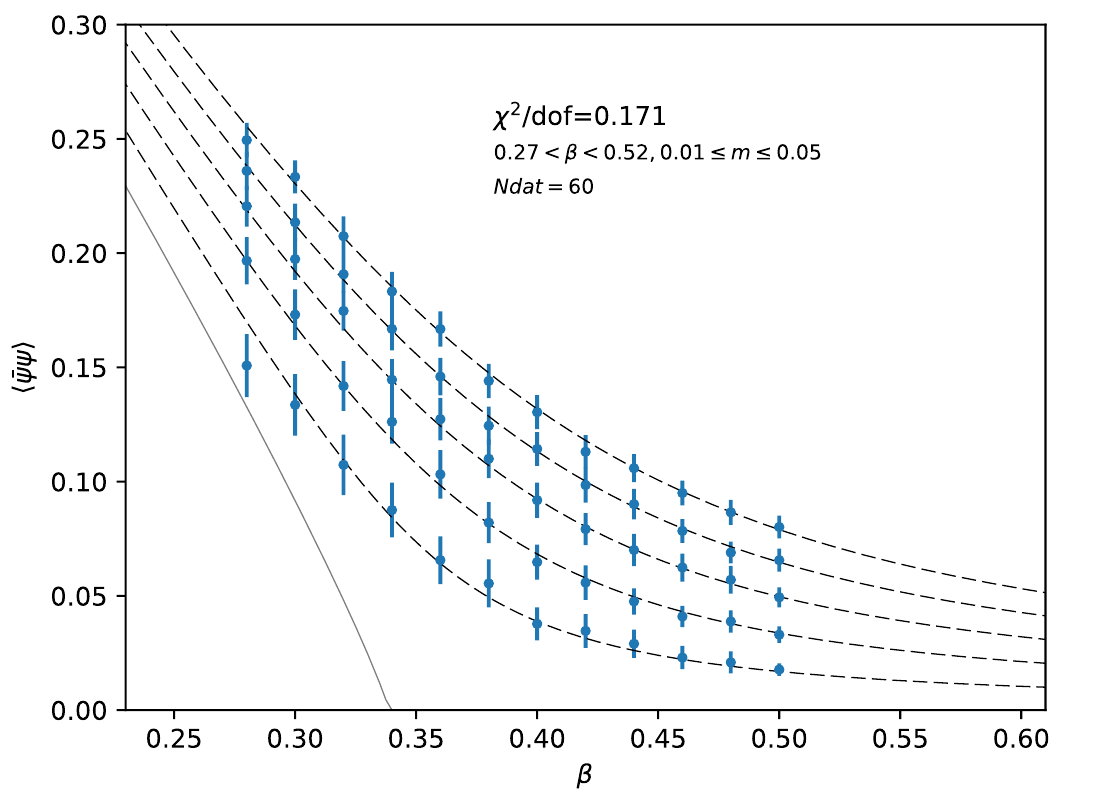}\label{fig:PQEoS}}
  \hspace*{4pt}
  \subfigure[Wilson kernel, $L_s({\rm sea})=30$HT, $L_s({\rm valence})=30$Z]
     {\includegraphics[width=2.8in]{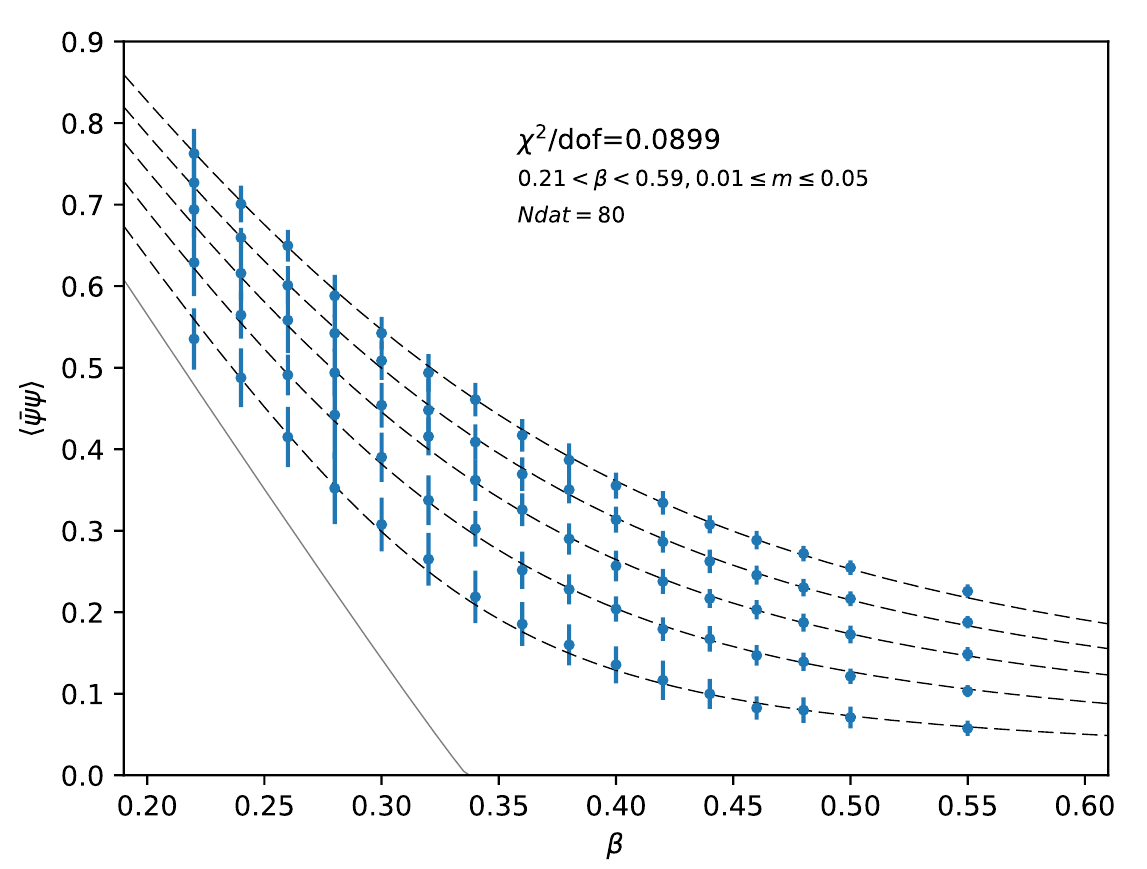}\label{fig:WilsonEoS}}
}
\caption{Thirring model equation of state on $12^3$}
\label{fig:EoS} 
\end{figure}
Since determination of the bilinear condensate on a finite system requires
$m\not=0$,
our approach to characterising the critical properties is to fit a
renormalisation group-inspired equation of state (EoS) to data collected in the
critical regime but with $m>0$:
\begin{equation}
m=A(g^{-2}-g_c^{-2})\langle\bar\psi\psi\rangle^{\delta-1/\beta}+B\langle\bar\psi\psi\rangle^\delta.
\label{eq:EoS}
\end{equation}
Data from a partially-quenched approach using the Shamir kernel is shown in
Fig.~\ref{fig:PQEoS} and from the Wilson kernel with Zolotarev approximation to sgn in
the valence sector in
Fig.~\ref{fig:WilsonEoS}.
\begin{table}
\begin{tabular*}{\textwidth}{@{}l@{\extracolsep{\fill}}llllll}
\hline
&Shamir PQ $12^3$ &  Wilson PQ $12^3$ &  Shamir $16^3$  & staggered $16^3$ &
staggered FSS \\
& $L_s({\rm v})=300$HT &  $L_s({\rm v})=30$Z &  \cite{Hands:2020itv}  &
\cite{DelDebbio:1997dv}  & \cite{Chandrasekharan:2011mn} \\
\hline
$ag_c^{-2}$ & 0.339(24) & 0.336(33) & 0.283(1) & - & -  \\
$\beta$ & 0.89(26) & 1.04(29)  & 0.320(5) & 0.57(2) & 0.70(1)\\
$\delta$ & 2.07(40)  & 2.08(33)  & 4.17(5) & 2.75(9) & 2.63(2)  \\ 
$\nu$ & 0.91(28)  & 1.1(3) & 0.55(1)  & 0.71(3) & 0.85(1) \\
$\eta$ & 0.96(18) & 0.95(15)  & 0.16(1)  & 0.60(4)  & 0.65(1)  \\
\hline
\end{tabular*}
\caption{Critical parameter fits}\label{tab:critical}
\end{table}
The fitted critical coupling $ag_c^{-2}$ and exponents $\beta,\delta$ are
tabulated in Table~\ref{tab:critical}, along with further exponents $\nu,\eta$
estimated from hyperscaling. For comparison results from the earlier
study~\cite{Hands:2020itv} based on extrapolating Shamir kernel data to
$L_s\to\infty$ using the {\em Ansatz} (\ref{eq:Ansatz}), as well as two
complementary studies of the Thirring model formulated with staggered lattice
fermions, one using the HMC algorihm on fixed volume fitting the EoS
(\ref{eq:EoS})~\cite{DelDebbio:1997dv}, and one using the fermion bag algorithm
to perform a finite volume scaling analysis~\cite{Chandrasekharan:2011mn}, are
also shown.

Since Shamir and Wilson kernels are in effect two distinct
regularisations of the Thirring model, we expect the derived critical exponents to
coincide.
While larger volumes and more statistics are needed to make definitive
conclusions, the compatibility of the results from the two new approaches is
encouraging, and consistent with universal scaling at a critical point for
$N=1$.  The new results are also clearly incompatible with previous published results 
\cite{Hands:2020itv}, suggesting that the exponential extrapolation
(\ref{eq:Ansatz}) is not controlling the large-$L_s$ limit  at accessible
values of $L_s$; in particular the approach seems to under-estimate the
critical $g_c^{-2}$. Finally, it is worth remarking that the Thirring model
defined using DWF yields distinct critical properties to those of the staggered
fermion model, corroborating the claim that at a strongly-coupled continuum
limit  naive taste
symmetry recovery cannot be assumed. Rather, it may well be that a distinct
interacting fermion theory based on the U($N)\otimes$U($N$) symmetry of
K\"ahler-Dirac fermions may exist, as outlined
in \cite{Hands:2021mrg}.

\section{Acknowledgements}
This work used the DiRAC Data Intensive service (CSD3) at the University of
Cambridge, managed by the University of Cambridge University Information
Services on behalf of the STFC DiRAC HPC Facility (www.dirac.ac.uk). The DiRAC
component of CSD3 at Cambridge was funded by BEIS, UKRI and STFC capital funding
and STFC operations grants. DiRAC is part of the UKRI Digital Research
Infrastructure.  Further work was performed on the Sunbird facility of
Supercomputing Wales. The work of JW was supported by an EPSRC studentship, and
of SH by STFC Consolidated Grant  ST/ST000813/1.

\end{document}